\documentclass[12pt,dvips]{article}

\usepackage{rotating}
\usepackage{axodraw}
\usepackage{epsfig}
\usepackage{color}
\usepackage{hhline}
\usepackage{graphicx}

\newcommand{\s}{\\ \vspace*{-3.5mm}}
\newcommand{\imag}{\Im {\rm m}}
\newcommand{\real}{\Re {\rm e}}
\newcommand{\rb}[2]{\raisebox{#1}[-#1]{#2}}
\newcommand{\sts}{\scriptstyle}

\begin{document}

\begin{titlepage}
\thispagestyle{empty}
\setcounter{page}{0}

\newcommand{\eeHZ}{$e^+e^- \to ZH$}
\newcommand{\epem}{$e^+e^-$}

\begin{flushright}
  CERN--TH/2002--231\\
  DESY 02-150\\
  PM/02-26\\
  hep--ph/0210077\\
\end{flushright}

\vspace{1cm}

\renewcommand{\thefootnote}{\fnsymbol{footnote}}

\begin{center}
{\Large \bf Identifying the Higgs Spin and Parity \\
            \vspace{3mm}
            in Decays to \boldmath{$Z$} Pairs }\\[1.5cm]
{\large S.Y. Choi$^{1}$, D.J. Miller$^2$, M.M. M\"uhlleitner$^{3}$ 
         and P.M.~Zerwas$^{4}$ }\\[1.cm]
{\it $^1$Chonbuk National University, Chonju 561-756, Korea\\[1mm]
     $^2$Theory Division, CERN, Geneva, Switzerland\\[1mm]
     $^3$Universit\'e de Montpellier II, F--34095 Montpellier 
         Cedex 5, France\\[1mm]
     $^4$Deutsches Elektronen--Synchrotron DESY, D--22603 Hamburg, Germany}
\end{center}

\renewcommand{\thefootnote}{\arabic{footnote}}
\vspace{4cm}

\begin{abstract}
\noindent
Higgs decays to $Z$ boson pairs may be exploited to determine spin and
parity of the Higgs boson, a method complementary to spin--parity
measurements in Higgs-strahlung. For a Higgs mass above the on-shell
$ZZ$ decay threshold, a model-{\linebreak}independent analysis can be performed,
but only by making use of additional angular correlation effects in
gluon-gluon fusion at the LHC and $\gamma \gamma$ fusion at linear
colliders. In the intermediate mass range, in which the Higgs boson
decays into pairs of real and virtual $Z$ bosons, threshold effects
and angular correlations, parallel to Higgs-strahlung, may be adopted
to determine spin and parity, though high event rates will be required
for the analysis in practice.
\end{abstract}

\end{titlepage}

\section{Introduction}
\noindent
The Higgs boson in the Standard Model must necessarily be a scalar
particle, assigned the external quantum numbers ${\cal J}^{\cal
PC}=0^{++}$; extended models such as ${\cal CP}$--invariant
supersymmetric theories also contain these pure scalar states. The
assignment of the quantum numbers invites investigating experimental
opportunities to identify spin and parity of the Higgs state at future
high-energy colliders. The determination of the parity and the parity
mixing of spinless Higgs bosons have been extensively investigated in
Refs.\cite{Kramer}-\cite{Bower}. The model--independent
identification of spin and parity of the Higgs particle has recently
been demonstrated for Higgs--strahlung, $e^+e^- \rightarrow ZH$, in
Ref.\cite{MCEMZ}, and experimental simulations have been performed in
Ref.\cite{Lohmann}. The rise of the excitation curve near the
threshold combined with angular distributions render the spin-parity
analysis of the Higgs boson unambiguous in this channel. \s

In the present note we study methods by which the spinless nature and the
positive parity of the Higgs boson can be identified through the decay 
process
\begin{eqnarray}
H \rightarrow ZZ\rightarrow (f_1\bar{f}_1)\, (f_2\bar{f}_2)
\label{eq:HZZ}
\end{eqnarray}
This process includes clean $\mu^+\mu^-$ and $e^+e^-$ decay channels for 
isolating the signal from the background and allowing a complete
reconstruction of the kinematical configuration with good precision
\cite{AT-TDR,CMS-TDR,Hohl}. While the dominant decay mode for Higgs masses 
below $\sim 140$ GeV is the $b\bar{b}$ decay channel, the $ZZ$ mode, 
one of the vector bosons being virtual below the threshold for two 
real $Z$ bosons, becomes leading for higher masses next to the 
$WW$ decay channel. \s
 
Higgs decays to $Z$ bosons can provide us with a clear picture of these
external quantum numbers 
for Higgs masses above the $ZZ$ threshold, if auxiliary angular distributions 
are included that are generated in specific production mechanism
such as gluon fusion at the LHC and $\gamma \gamma$ fusion at
linear colliders. Below the mass range for on-shell $ZZ$ decays, threshold 
analyses combined with angular correlations in $Z^*Z$ decays [with one of the 
electroweak bosons, $Z^*$, being virtual] may be exploited in analogy to
Higgs-strahlung at $e^+e^-$ linear colliders. The picture is theoretically 
transparent in this mass range but low rates and large backgrounds render 
this $Z$ decay channel quite difficult for the analysis of spin and parity 
of the Higgs particle.\s

\section{Heavy Higgs Bosons}
\noindent
Above the on-shell $ZZ$ threshold, the partial width for Higgs decays 
into $Z$ boson pairs is given in the Standard Model by the expression
\begin{eqnarray}
\Gamma(H\rightarrow ZZ) = \frac{\sqrt{2} G_F}{16\pi} M^3_H 
     \left(1-4x+12x^2\right)\, \beta
\end{eqnarray}
where $x=M^2_Z/M^2_H$, and $\beta=\sqrt{1-4M_Z^2/M_H^2}$
is the velocity of the $Z$ bosons in the Higgs rest frame. 
For large Higgs masses, the $Z$ bosons are longitudinally polarized 
according to the equivalence principle, while the longitudinal and transverse
polarization states are populated democratically near the threshold. \s

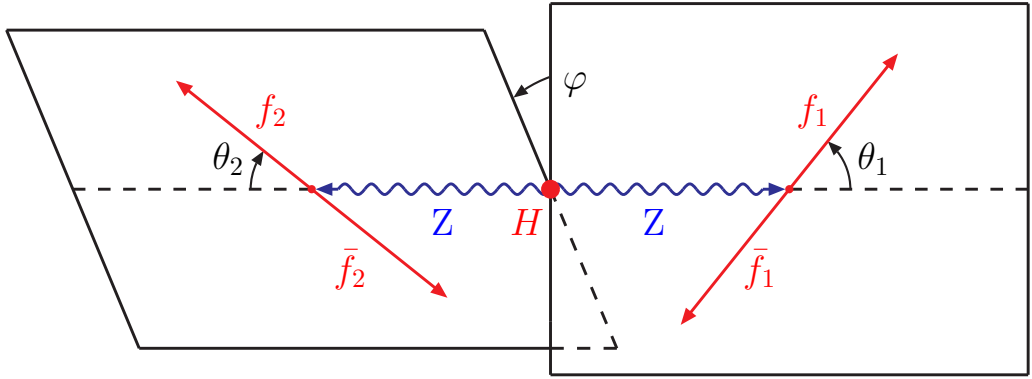
\begin{figure}[t!]
\begin{center}
\begin{picture}(550,200)(0,0)
\SetWidth{1.1}
%1st diagram 
\SetColor{Blue}
\Photon(230,100)(310,100){2}{8}
\LongArrow(310,100)(317,100)
\Photon(230,100)(150,100){2}{8}
\LongArrow(150,100)(143,100)
\SetColor{Black}
\DashLine(320,100)(410,100){5}
\DashLine(140,100)(50,100){5}
\Line(230,170)(230,100)
\Line(230,50)(230,30)
%\DashLine(230,100)(230,50){5}
\Line(230,100)(230,50)
\Line(230,170)(410,170)
\Line(230,30)(410,30)
\Line(410,170)(410,30)
\Line(205,160)(230,100)
\DashLine(230,100)(255,40){5}
%\Line(230,100)(250,50)
\Line(25,160)(75,40)
\Line(25,160)(205,160)
\Line(75,40)(230,40)
\DashLine(230,40)(255,40){5}
%\Line(230,50)(250,50)
\SetWidth{0.8}
\LongArrowArc(320,100)(23,0,51.34)
%\CArc(320,100)(20,0,51.34)
\LongArrowArcn(140,100)(23,180,141.34)
%\CArc(140,100)(20,141.34,180)
%\LongArrowArc(230,100)(40,90,111.8)
\LongArrowArc(245,100)(45,110,130)
%\CArc(230,100)(38,90,111.8)
%\CArc(230,100)(35,90,111.8)
\SetWidth{1.1}
\Text(222,88)[]{\large\color{red} $H$}
\Text(270,88)[]{\large\color{blue} Z}
\Text(190,88)[]{\large\color{blue} Z}
\Text(330,130)[c]{\large\color{red} $f_1$}
\Text(310,70)[c]{\large\color{red} $\bar{f}_1$}
\Text(125,130)[c]{\large\color{red} $f_2$}
\Text(155,70)[c]{\large\color{red} $\bar{f}_2$}
\Text(353,112)[c]{\large\color{black} $\theta_1$}
\Text(109,112)[c]{\large\color{black} $\theta_2$}
\Text(240,140)[c]{\large\color{black} $\varphi$}
\SetColor{Red}
\LongArrow(140,100)(90,140)
\LongArrow(140,100)(190,60)
\LongArrow(320,100)(360,150)
\LongArrow(320,100)(280,50)
\CCirc(230,100){3}{Red}{Red}
\CCirc(320,100){1}{Red}{Red}
\CCirc(140,100){1}{Red}{Red}
\end{picture}
\end{center}
\caption{{\it The definition of the polar angles ${\theta_i}$ 
($i=1,2$) and the azimuthal angle $\varphi$ for the sequential decay
$H \rightarrow Z^{(*)} Z \rightarrow (f_1\bar{f}_1) (f_2\bar{f}_2)$ in
the rest frame of the Higgs particle.}}
\label{fig:angles}
\end{figure}

The characteristic observables for measuring spin and parity of the
Higgs boson are the angular distributions of the final-state fermions
in the decays $Z \rightarrow f\bar{f}$, encoding the helicities of the
$Z$ states.  The combined polar and azimuthal angular distributions
are presented for the Standard Model in the Appendix. \s

%Denoting the polar angles of the fermions $f_1, f_2$ in the rest
%fames of the $Z$ bosons by $\theta_1$ and $\theta_2$, and the
%azimuthal angle between the fermion pairs by $\phi$, [see
%Fig.\ref{fig:angles}], the differential distribution in
%$\cos\theta_1$, $cos\theta_2$ and $\phi$ is predicted by the Standard
%Model to be 
%
%\begin{eqnarray}
%\frac{d\Gamma_H}{d dc_{\theta_1} dc_{\theta_2} d \phi}& \sim&
%   s_{\theta_1}^2s_{\theta_2}^2 - \frac{1}{2 \gamma^2(1+\beta^2)} 
%   s_{2\theta_1} s_{2\theta_2} c_{\phi} \nonumber \\
%&& \hspace{-1.5cm} 
%   +\, \frac{1}{2 \gamma^4(1+\beta^2)^2}
%    \left[(1+c_{\theta_1}^2)(1+c_{\theta_2}^2)
%   +s_{\theta_1}^2 s_{\theta_2}^2 c_{2\phi} \right] \nonumber \\
%&& \hspace{-1.5cm} 
%   +\, \frac{2 v_1 a_1}{v^2_1+a^2_1}\frac{2 v_2 a_2}{v^2_2+a^2_2}
%    \frac{2}{\gamma^2(1+\beta^2)} \left[ - s_{\theta_1} s_{\theta_2} c_{\phi}
%   + \frac{1}{\gamma^2(1+\beta^2)} c_{\theta_1} c_{\theta_2} \right]
%\end{eqnarray}
%
Polar and azimuthal angular distributions give independent access to
spin and parity of the Higgs boson.  Denoting the polar angles of the
fermions $f_1, f_2$ in the rest frames of the $Z$ bosons by $\theta_1$
and $\theta_2$, and the azimuthal angle between the planes of the
fermion pairs by $\varphi$, [see Fig.\ref{fig:angles}], the differential distribution
in $\cos\theta_1$, $\cos\theta_2$ is predicted by the Standard Model
to be
\begin{eqnarray}
\frac{d\Gamma_H}{d \cos{\theta_1} d \cos{\theta_2}} &\sim&
   \sin^2{\theta_1} \sin^2{\theta_2} \nonumber \\
    && \,+\frac{1}{2 \gamma^4 (1+\beta^2)^2}\left[
    (1+\cos^2{\theta_1})(1+\cos^2{\theta_2}) 
   +4\, \eta_1 \, \eta_2
    \,\cos{\theta_1} \cos{\theta_2}\right]
\label{eq:dist_polar}
\end{eqnarray}
while the corresponding distribution with respect to the azimuthal
angle $\varphi$ is
\begin{eqnarray}
\frac{d\Gamma_H}{d \varphi} &\sim & 1
 -\eta_1\eta_2 \, \frac{1}{2}{\left(\frac{3\pi}{4}\right)}^2
  \frac{\gamma^2(1+\beta^2)}{
   \gamma^4 (1+\beta^2)^2 +2}\cos\varphi
 +\frac{1}{2}\frac{1}{\gamma^4 (1+\beta^2)^2 
    +2}\cos2\varphi
\label{eq:dist_azimuth}
\end{eqnarray}
%
%
%\begin{eqnarray}
%\frac{d\Gamma_H}{d \varphi} &\sim&
%1-\frac{1}{2 \gamma^2(1+\beta^2)} \cos\varphi
%+\frac{2}{\gamma^4(1+\beta^2)^2} \left(1+\frac{1}{4} 
%\cos{2\varphi} \right) \nonumber \\
%&&+\, \eta_1 \, \eta_2
%  \, \frac{9 \pi^2}{32 \, \gamma^2(1+\beta^2)} \cos{\varphi}
%\label{eq:dist_azimuth}
%\end{eqnarray}
%
\noindent
where $\eta_i = 2v_i a_i/(v_i^2+a_i^2)$ is the polarization degree
with the electroweak charges \linebreak \mbox{$v_i=2 I_{3i}- 4 e_i
\sin^2\theta_W$} and $a_i= 2 I_{3i}$ of the fermion $f_i$; and $\gamma
= 1/\sqrt{1-\beta^2}$ is the Lorentz-boost factor of the $Z$ bosons.
For large Higgs masses, the longitudinal $Z$ polarization is reflected
in the asymptotic behaviour of the double differential distribution,
approaching $\sim \sin^2\theta_1\,\sin^2\theta_2$ in this limit.
Also any $\varphi$ dependence disappears in this limit.  The $\varphi$
distribution has been analyzed in a recent experimental simulation as
a tool to shed light on Higgs spin measurements at the LHC,
Ref.~\cite{Hohl}.

As a discriminant, the two distributions (\ref{eq:dist_polar}) and 
(\ref{eq:dist_azimuth}) can readily be 
confronted with the decay distributions of a pseudo-scalar particle into 
two $Z$ bosons carrying the momenta $k_1$ and $k_2$.
While the scalar decay amplitude can be expressed as 
a scalar product of the two $Z$ polarization vectors, $A_{+} =
\varepsilon^*_{1} \cdot \varepsilon^*_{2}$, dominated by the large 
longitudinal
wave functions, the pseudo-scalar decay amplitude, 
$ A_{-} = {\rm {det}}[k_1,k_2,\varepsilon^*_{1},\varepsilon^*_{2}] \sim 
{\stackrel{\rightarrow}{k_1}} \cdot ({\stackrel{\rightarrow}{\epsilon^*_1}}
  \times {\stackrel{\rightarrow}{\epsilon^*_2}})$, 
is non-vanishing only for transverse $Z$ polarization which gives
rise to the following angular distributions, independent of the
Higgs-mass value:
\begin{eqnarray}
\frac{d\Gamma}{d\cos{\theta_1} d\cos{\theta_2}} \sim
(1+\cos^2{\theta_1})(1+\cos^2{\theta_2}) 
 + 4\, \eta_1 \eta_2 
 \cos{\theta_1} \cos{\theta_2}
\end{eqnarray} 
\noindent
and
\begin{eqnarray}
\frac{d\Gamma}{d\varphi} \sim {1 - {\frac{1}{4}} \cos{2 \varphi}}
\end{eqnarray}

\begin{figure}[htb!]
\begin{center}
\includegraphics[clip=true,trim=5 5 5 5,width=11cm]{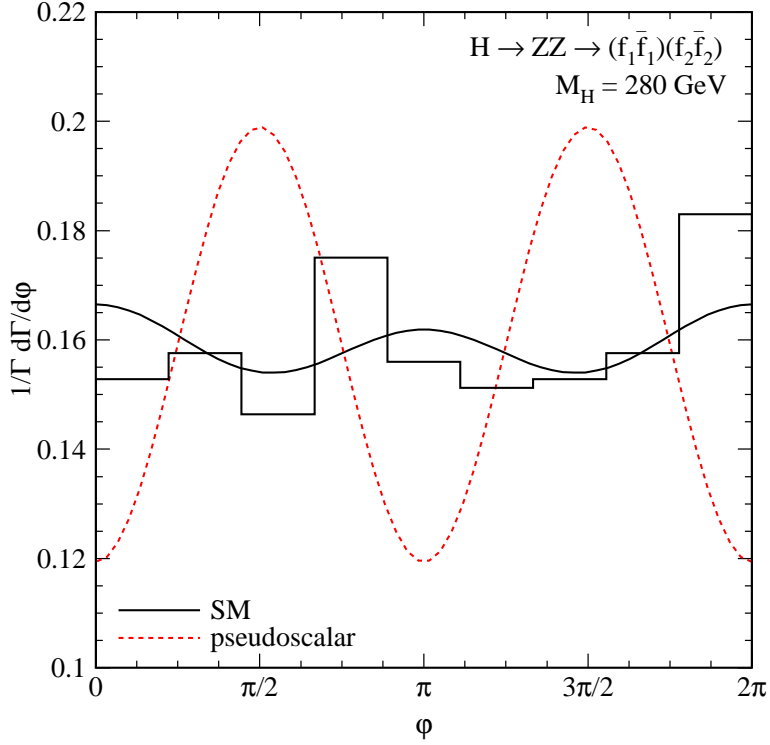}
\end{center}
\caption{{\it The azimuthal distributions, $d\Gamma/d\varphi$, 
for the Standard Model Higgs boson and a pseudoscalar boson,
with a Higgs mass of $280$ GeV. The histogram for the Standard Model
shows the expected result from $900$ signal events corresponding to 
an integrated luminosity of $\int {\cal L}\, dt = 300\, {\rm fb}^{-1}$ 
at LHC 
[with efficiencies and cuts included according to the
experimental simulation Ref.\cite{Hohl}]. The curves show the exact
theoretical dependences for the scalar and pseudoscalar, appropriately
normalised.}}
\label{fig:phi} 
\end{figure}

\noindent
The two distributions for negative-parity decays are distinctly different 
from the positive-parity form predicted by the Standard Model. This is shown 
for a Higgs mass $M_H = 280 \;$GeV in Fig.\ref{fig:phi} for the azimuthal 
distributions. The predictions will be distorted by experimental
cuts which however can be corrected for as shown in Ref.\cite{Hohl}. 
Moreover, the accuracy will improve significantly with rising statistics 
beyond the integrated luminosity adopted in the figure. \s

\noindent

This result can systematically be generalized to arbitrary spin and parity 
assignments of the decaying particle. The helicity
formalism is the most convenient theoretical
tool for performing this analysis. Denoting the basic helicity
amplitude~\cite{Helicity} for arbitrary $H$ spin-${\cal J}$ by 
\begin{eqnarray}
\langle Z(\lambda_1) Z(\lambda_2) | H_{\cal J}(m) \rangle = 
  \frac{g_{_W} M_Z}{\cos
  \theta_W} \, {\cal T}_{\lambda_1\lambda_2}\,
  d^{\cal J}_{m,\,\lambda_1-\lambda_2}(\Theta)\,
  {\rm e}^{ - i( m-\lambda_1+\lambda_2)\,\Phi} 
\label{eq:vert} 
\end{eqnarray}
the reduced vertex ${\cal T}_{\lambda_1 \lambda_2}$ depends only on
the helicities of the two real $Z$ bosons,
but it is independent of the $H$ spin component $m$ along the polarization
axis of the decaying particle. This axis is defined by the polar and 
azimuthal angles, $\Theta$ and $\Phi$, in the coordinate system in which
the momentum of the $Z$ boson decaying to $f_1 \bar{f_1}$ 
points to the positive $z$--axis and the $f_1$ momentum defines 
the $xz$ plane with the $x$-component taken positive, cf. 
Fig.\ref{fig:angles}. The standard coupling is split off explicitly. \s

The normality of the Higgs state, $n_H = (-1)^{\cal J} \,{\cal P}$, 
connects the helicity amplitudes under parity 
transformations. If the interactions determining the vertex (\ref{eq:vert}) 
are ${\cal P}$ invariant, equivalent to ${\cal CP}$ invariance in this 
specific case, the reduced vertices are related,
\begin{eqnarray}
{\cal T}_{ \lambda_1 \lambda_2} = n_H \, {\cal T}_{-\lambda_1 \, -\lambda_2} 
\label{eq:norm} 
\end{eqnarray}
Above the threshold for two real $Z$ bosons, the helicity amplitudes are
restricted further by Bose symmetry as
\begin{eqnarray}
{\cal T}_{\lambda_1\lambda_2}=(-1)^{\cal J}\, {\cal T}_{\lambda_2\lambda_1}
\label{eq:Bose}
\end{eqnarray}
independently of the parity of the decaying particle. \s

For a ${\cal CP}$ invariant theory the polar--angle 
distributions can be written in the form 
\begin{eqnarray}
\frac{d\Gamma}{d\cos{\theta_1} d\cos{\theta_2}} &\sim & 
   \sin^2{\theta_1}\sin^2{\theta_2}\, |{\cal T}_{00}|^2
     +\frac{1}{2}(1+\cos^2{\theta_1})(1+\cos^2{\theta_2})
      \left[|{\cal T}_{11}|^2+|{\cal T}_{1,-1}|^2\right]\nonumber\\
   &&\ \ \ \ +(1+\cos^2{\theta_1})\, \sin^2{\theta_2}\,|{\cal T}_{10}|^2 
         +\sin^2{\theta_1}\,(1+\cos^2{\theta_2})\,|{\cal T}_{01}|^2 \nonumber\\
   &&\ \ \ \ + 2\,\eta_1 \eta_2
           \cos{\theta_1} \cos{\theta_2}
             \left[|{\cal T}_{11}|^2-|{\cal T}_{1,-1}|^2\right]
\label{eq:ddhel1} 
\end{eqnarray}
while the general azimuthal angular distribution reads 
\begin{eqnarray}
\frac{d\Gamma}{d\varphi} &\sim& 
  |T_{11}|^2+|T_{10}|^2+|T_{1,-1}|^2 +|T_{01}|^2 +
            |T_{00}|^2/2\nonumber\\
 &&\ \ \ \ +\eta_1\eta_2{\left(\frac{3\pi}{8}\right)}^2
            \real(T_{11}T^*_{00}+T_{10}T^*_{0,-1})\cos\varphi
	   +\frac{1}{4}\real(T_{11}T^*_{-1,-1})\cos 2\varphi
\label{eq:ddhel2} 
\end{eqnarray}

\noindent
The helicity amplitudes of the decay $H \rightarrow ZZ$ in the Standard
Model are given by
\begin{eqnarray}
{\cal T}_{00} = M^2_H/(2M_Z^2) - 1, \quad
{\cal T}_{11} = -1, \quad
{\cal T}_{10}={\cal T}_{01}={\cal T}_{1,-1}\, =\, 0  
\end{eqnarray}
and the Higgs boson carries even normality: $n_H =+1$. \s 

The most general $HZZ$ vertex  
is given by the expression
\begin{eqnarray}
{\cal J} = \frac{g_W M_Z}{\cos \theta_W} \, 
   T_{\mu \nu \beta_1 ... \beta_{\cal J}} \, \varepsilon^*(Z_1)^{\mu} \,
   \varepsilon^*(Z_2)^{\nu} \,
   \varepsilon(H)^{\beta_1 ... \beta_{\cal J}} 
\label{eq:cur} 
\end{eqnarray}
While $\varepsilon^{\mu}$ and $\varepsilon^{\nu}$ are the usual
spin--1 polarization vectors, the spin--${\cal J}$ polarization tensor
$\varepsilon^{\beta_1 ...  \beta_{\cal J}}$ of the state $H$ has the
notable properties of being symmetric, traceless and orthogonal to the
4-momentum of the Higgs boson $p^{\beta_i}$, and it can be constructed
from products of suitably chosen polarization vectors. $T_{\mu \nu
\beta_1 ... \beta_{\cal J}}$ is normalized such that $T_{\mu \,
\nu}=g_{\mu \, \nu}$ in the Standard Model. Moreover, with the 
assumption of massless leptons in the final state, $T_{\mu \nu \beta_1
... \beta_{\cal J}}$ is transverse due to the conservation of the
lepton currents, strongly constraining the form of the
tensor\footnote{The most general tensor couplings of the $HZZ$ vertex
for Higgs particles of spin $\leq 2$ are listed in
Table.\ref{tab:ff}. }. \s

\begin{table}[ht]
\centering
\begin{tabular}{||c||l|l|c||} \hhline{|t:=:t:===:t|}
${\cal J^P}$&\multicolumn{1}{c|}{\small $HZ^*Z$ Coupling}
& \multicolumn{1}{c|}{\small Helicity Amplitudes} & \small Threshold
\\\hhline{|:=:b:===:|} 
\multicolumn{4}{||c||}{\small Even Normality $n_H=+$} \\\hhline{|:=:t:===:|}
&&$\sts {\cal T}_{00} =  
(2a_1(M_H^2-M_*^2-M_Z^2) + a_2 M_H^4 \beta^2) /(4M_*M_Z)$ & $\sts 1$ \\
\rb{1.5ex}{$0^+$}&
\rb{1.5ex}{$\sts \phantom{+}a_1\, g^{\mu\nu}+a_2\, p^{\mu}p^{\nu}$}&
$\sts {\cal T}_{11} = -a_1$& 
$\sts 1$ \\\hhline{||-||---||}
&&
$\sts {\cal T}_{00}= \beta \, [ -2b_1 (M_H^2-M_*^2-M_Z^2) -b_2(M_H^2-M_Z^2+M_*^2)$&\\
&&
$\sts \phantom{{\cal T}_{00}= \beta \, [ }
+b_3(M_H^2-M_*^2+M_Z^2)-b_4 M_H^4 \beta^2 ]M_H/(4\,M_* M_Z)$&
\rb{1.5ex}{$\sts \beta$} \\
$1^-$&
\rb{1.5ex}{$\sts \phantom{+}b_1 g^{\mu\nu}k^{\beta}+b_2g^{\mu\beta}p^{\nu}$}&
$\sts {\cal T}_{01}= \beta\,b_3 M_H^2/(2M_*) $ & $\sts \beta$ \\
&\rb{1.5ex}{$\sts +b_3g^{\nu\beta}p^{\mu}+b_4p^{\mu}p^{\nu}k^{\beta}$}&
$\sts {\cal T}_{10}=-\beta\,b_2 M_H^2/(2M_Z)$ & $\sts \beta$\\
&&
$\sts {\cal T}_{11}=\beta\,b_1M_H$ & $\sts \beta$ \\\hhline{||-||---||}
&&
$\sts {\cal T}_{00}=\big\{-c_1\,(M_H^4-(M_Z^2-M_*^2)^2)/M_H^2 $ &\\
&
$\sts \phantom{+}c_1\, (g^{\mu\beta_1}g^{\nu\beta_2}+g^{\mu\beta_2}g^{\nu\beta_1})$&
$\sts \phantom{{\cal T}_{00}=\big\{}+M_H^2 \beta^2[c_2\,
(M_H^2-M_Z^2-M_*^2)+c_3\,(M_H^2-M_Z^2+M_*^2)$ &$\sts 1$\\
&
$\sts +c_2\,g^{\mu\nu}\,k^{\beta_1}k^{\beta_2}$&
$\sts \phantom{{\cal T}_{00}=\big\{}-c_4\,(M_H^2-M_*^2+M_Z^2) \,]
+\frac{1}{2}c_5\,M_H^{6}\beta^4 \big\}/(\sqrt{6}M_ZM_*)$&\\
$2^+$&
$\sts +c_3\,(g^{\mu \beta_1}k^{\beta_2}+g^{\mu\beta_2}k^{\beta_1})\,p^{\nu}$&
$\sts {\cal T}_{01}=(-c_1(M_H^2-M_Z^2+M_*^2)-c_4\,M_H^4 \beta^2)/(\sqrt{2}M_*M_H)$&
$\sts 1$\\
&
$\sts +c_4\,(g^{\nu\beta_1} k^{\beta_2}+g^{\nu\beta_2} k^{\beta_1}) \, p^{\mu}$&
$\sts {\cal T}_{10}=(-c_1(M_H^2-M_*^2+M_Z^2)+c_3\,M_H^4 \beta^2)/(\sqrt{2}M_ZM_H)$&
$\sts 1$\\
&
$\sts +c_5\,p^{\mu}p^{\nu}k^{\beta_1}k^{\beta_2}$&
$\sts {\cal T}_{11}=-\sqrt{2/3}\,(c_1+c_2M_H^2\beta^2)$ &
$\sts 1$\\
&&
$\sts {\cal T}_{1,-1}=-2\, c_1$&$\sts 1$\\\hhline{|:=:b:===:|} 
\multicolumn{4}{||c||}{\small Odd Normality $n_H=-$} \\\hhline{|:=:t:===:|}
&&
$\sts{\cal T}_{00}=0$&\\
\rb{1.5ex}{$0^-$}&
\rb{1.5ex}{$\sts \phantom{+}a_1\,\epsilon^{\mu\nu\rho\sigma}p_{\rho}k_{\sigma}$}&
$\sts{\cal T}_{11}=i\,\beta\,M_H^2\,a_1$&$\sts \beta$\\\hhline{||-||---||}
&&
$\sts{\cal T}_{00}=0$&\\
&\rb{0ex}{$\sts\phantom{+}b_1\,\epsilon^{\mu\nu\beta\rho}p_{\rho}$}&
$\sts{\cal T}_{01}=i\,(b_1\,(M_Z^2-M_H^2-M_*^2)+b_2\,(M_H^2-M_Z^2-3M_*^2)$&\\
&&
$\sts \phantom{{\cal T}_{01}=i\,(} +b_3\,M_H^4 \beta^2)/(2M_*)$&
\rb{1.5ex}{$\sts 1$}\\
\rb{1.5ex}{$1^+$}&
\rb{3ex}{$\sts +b_2\,\epsilon^{\mu\nu\beta\rho}k_{\rho}$}&
$\sts{\cal T}_{10}=i\,(b_1\,(M_*^2-M_H^2-M_Z^2)-b_2\,(M_H^2-M_*^2-3M_Z^2)$&\\
&
\rb{3ex}{$\sts+b_3\,(\epsilon^{\mu\beta\rho\sigma}p^{\nu}$}&
$\sts \phantom{{\cal T}_{01}=i\,(} +b_3\,M_H^4 \beta^2)/(2M_Z)$&
\rb{1.5ex}{$\sts 1$}\\
&
\rb{3ex}{$\sts\phantom{+b_3\,(+}
+\,\epsilon^{\nu\beta\rho\sigma}p^{\mu} )\, p_{\rho}k_{\sigma}$}&
$\sts{\cal T}_{11}=i\,(-b_1\,M_H^2+b_2\,(M_Z^2-M_*^2))/M_H$&
$\sts 1$\\\hhline{||-||---||}
&&
$\sts{\cal T}_{00}=0$&\\
&
\rb{1.5ex}{$\sts\phantom{+}c_1\,\epsilon^{\mu\nu\beta_1\rho}p_{\rho}k^{\beta_2}$} &
$\sts{\cal T}_{01}=i\,\beta\,(c_1\,(M_H^2+M_*^2-M_Z^2)-c_2\,(M_H^2-M_Z^2-3M_*^2)$&\\
&
\rb{1.5ex}{$\sts+c_2\,\epsilon^{\mu\nu\beta_1\rho}k_{\rho}k^{\beta_2}$}&
$\sts \phantom{{\cal T}_{01}=i\,\beta\,(} -c_3M_H^4 \beta^2)M_H/(\sqrt{2}M_*)$&
\rb{1.5ex}{$\sts \beta$}\\
$2^-$&
\rb{1.5ex}{$\sts+c_3\,(\epsilon^{\mu\beta_1\rho\sigma}p^{\nu}$}&
$\sts{\cal T}_{10}=i\,\beta\,(c_1\,(M_H^2+M_Z^2-M_*^2)+c_2\,(M_H^2-M_*^2-3M_Z^2)$&\\
&
\rb{1.5ex}{$\sts\phantom{+c_3\,(+} 
+\,\epsilon^{\nu\beta_1\rho\sigma}p^{\mu})\,k^{\beta_2} p_{\rho}k_{\sigma}$}&
$\sts \phantom{{\cal T}_{10}=i\,\beta\,(} -c_3M_H^4 \beta^2)M_H/(\sqrt{2}M_Z)$&
\rb{1.5ex}{$\sts \beta$}\\
&
\rb{1.5ex}{$\sts+c_4 \,\epsilon^{\mu\nu\rho\sigma}
p_{\rho}k_{\sigma}k^{\beta_1}k^{\beta_2}$}&
$\sts{\cal T}_{11}=i\,\beta\,2\sqrt{2/3}\,(c_1\,M_H^2+c_2\,(M_*^2-M_Z^2)+c_4\,M_H^4 \beta^2)$&
$\sts \beta$\\
&
\rb{1.5ex}{$\sts+ \beta_1 \leftrightarrow \beta_2$}& 
$\sts{\cal T}_{1,-1}=0$& \\\hhline{|b:=:b:===:b|}
\end{tabular}
\caption{\it The most general tensor couplings of the $HZ^*Z$ 
vertex and the corresponding helicity amplitudes for Higgs bosons of
spin~$\leq 2$.  Here $p=k_1+k_2$ and \mbox{$k=k_1-k_2$}, where $k_1$
and $k_2$ are the 4-momenta of the $Z^*$ and the $Z$ bosons
respectively.  For spin $\geq 3$, the helicity amplitudes rise 
$\sim \beta^{{\cal J}-2}$ and $\sim \beta^{{\cal J}-1}$ for even and 
odd normalities respectively.}
\label{tab:ff}
\end{table}
\noindent
\underline {\bf {Odd normality:}}\\
\noindent
When comparing with the prediction of the Standard Model, it is quite
easy to rule out all states for odd normality: ${\cal J}^{\cal P} =
0^{-}$, $1^{+}$, $2^{-}$, $3^{+}$, $\ldots$. Since the helicity
amplitude ${\cal T}_{00}$ must vanish by the relation (\ref{eq:norm})
for odd normality, the observation of a non-zero $\sim \,
\sin^2\theta_1\,\sin^2\theta_2$ correlation in Eq.(\ref{eq:ddhel1}) as
predicted by the Standard Model, eliminates all odd-normality
states. \s

\noindent
\underline {\bf {Even normality:}}\\
\noindent
In the chain of even-normality states ${\cal J}^{\cal P} = 1^-$, $2^+$,
$3^-$, $4^+$, $\ldots$, the odd-spin states $1^-$, $3^-$, $\ldots$, can easily
be excluded by observing the $\sin^2\theta_1\,\sin^2\theta_2$ correlation
induced by ${\cal T}_{00}$ in the Standard Model, but forbidden by 
Bose symmetry for even-normality odd-spin states. \s

Excluding even-normality even-spin states $2^+$, $4^+$, $\ldots$ is a much
more difficult task. In general, the vertex 
   (\ref{eq:vert}) for the higher even--${\cal J}$ Higgs state will lead to 
   four--fermion angular correlations different from those for 
   the spin--0 case. However, if the tensor $T_{\mu\nu\beta_1...
   \beta_{\cal J}}$ is of the form
   \begin{eqnarray}
   T_{\mu\nu\beta_1...\beta_{\cal J}}=\left[\, T^{{\cal J}=0}_{\mu\nu} 
   \right]\, k_{\beta_1...} k_{\beta_{\cal J}}
   \label{eq:special}
   \end{eqnarray}
[with $k = k_1-k_2$], the unpolarized 
higher even--${\cal J}$ state generates 
the same angular correlations of the $Z$ decay products as the spin--0 state. 
Thus, from final-state distributions alone, without exploiting non-trivial 
helicity information from the decaying state, a model-independent spin-parity
analysis cannot be carried out. \s

However, special production mechanisms such as gluon fusion $gg
\rightarrow H$ at LHC \cite{AT-TDR,CMS-TDR} and photon fusion $\gamma
\gamma \rightarrow H$ in the Compton mode of linear colliders
\cite{TESLA} can be successfully exploited to close the gap. \s
 
In the gluon fusion process $gg \rightarrow H$, which is the dominant
Higgs production process in the Standard Model at the LHC,
Refs.\cite{Glashow,GrauZ}, the produced states transport non-trivial
spin information.  The most general spin--${\cal J}$ tensor
$\Gamma^{\mu\nu\beta_1... \beta_{\cal J}}$ for the $ggH$
coupling\footnote{Large QCD radiative corrections \cite{GrauZ,Djou} to
Higgs production in gluon fusion are built up in the
infrared gluon region and they do not affect strongly the state of
spin.}, apart from trivial factors, is the direct product of the
tensor $\Gamma^{\mu \nu \beta_i\beta_j}_{(2)}$
\begin{eqnarray}
   \Gamma^{\mu\nu\beta_i\beta_j}_{(2)}
     =   a_1 g^{\mu\nu}_\perp q^{\beta_i} q^{\beta_j}
     +   a_2 (g^{\mu\beta_i}_\perp g^{\nu\beta_j}_\perp
           + g^{\mu\beta_j}_\perp g^{\nu\beta_i}_\perp)\,M_H^2
\end{eqnarray}
isomorphic with the spin-2 tensor, and direct products of the momentum
vectors \linebreak \mbox{$q^{\beta}=(q_1-q_2)^{\beta}$} of the two
gluon momenta $q_1$ and $q_2$, as required by the properties of the
spin-${\cal J}$ wave-function $\varepsilon^{\beta_1 ...\beta_{\cal
J}}$. Here, the metric tensors, $g^{\mu{\beta_i}}_\perp$ and $g^{\nu
\beta_j}_{\perp}$, are defined to be orthogonal to $q^\mu_1$ and
$q^\nu_2$, while the tensor $g^{\mu\nu}_\perp$ is orthogonal to both
$q^\mu_1$ and $q^\nu_2$.  This tensor also describes the spin-0 state
[\,while the spin-1 tensor vanishes as spin--1 states do not couple to
pairs of gluons or photons according to Yang's theorem].  Assuming the
$HZZ$ coupling to be of the form (\ref{eq:special}), the polar--angle
distribution for the process $gg\rightarrow H \rightarrow ZZ$ is given
by the differential cross section
\begin{eqnarray}
   \frac{d\sigma}{d\cos\Theta}\left[\,gg\rightarrow H\rightarrow ZZ\right]
   \sim  |a_1|^2 \left[P^0_{\cal J}(\cos\Theta)\right]^2
     + 12 |a_2|^2 \left[P^2_{\cal J}(\cos\Theta)\right]^2
\end{eqnarray}
where $\Theta$ is the polar angle between the momenta of a gluon and a
$Z$ boson in the $gg$ center--of--mass frame.  The two functions
${\cal P}^{2}_{\cal J}$ and ${\cal P}^{0}_{\cal J}$ are associated
Legendre functions with non-trivial $\cos\Theta$ dependence except for
${\cal J} = 0$, see Ref.\cite{Messiah}.  Therefore, the distribution
is isotropic only for a spin--0 Higgs particle, but it is
an--isotropic for all higher even--spin Higgs particles. Thus, the
zero--spin of the Higgs boson can be checked through the lack of the
polar (and azimuthal) angle correlations between the initial state and
final state particles in the combined process of production
$gg\rightarrow H$ and decay $H\rightarrow ZZ$. [The transition from
$gg$ to $\gamma
\gamma \rightarrow H \rightarrow ZZ$, cf. Ref.\cite{Jik} for the
Standard Model, follows the same pattern.] \s

\section{Intermediate Higgs-Mass Range}
\noindent
Rates for Higgs decays $H \rightarrow Z^*Z$ to a pair of virtual and real 
$Z$ bosons are suppressed by one power of the electroweak coupling,
so that only a limited sample of events can be exploited for 
detailed analyses beyond the search, see e.g. Refs.\cite{AT-TDR,CMS-TDR,Hohl}.
Nevertheless, we will summarize the essential points for measuring Higgs spin 
and parity in this intermediate mass range. The analysis runs parallel 
in all elements to the same task in Higgs-strahlung at $e^+e^-$ colliders 
-- just requiring the crossing of the virtual $Z$-boson line 
from the initial to the final state. \s

Below the threshold of two real $Z$ bosons, the Higgs particle can decay
into real and virtual $Z^*Z$ pairs. The partial decay width is given 
in the Standard Model by
\begin{eqnarray}
\Gamma(H\rightarrow Z^*Z)=\frac{3 G^2_F M^4_Z}{16\pi^3}\delta_Z M_H R(x),
\end{eqnarray}
where $\delta_Z=7/12-10\sin^2\theta_W/9+ 40\sin^4\theta_W/27$, and the 
expression for $R(x)$, 
\begin{eqnarray}
R(x) &=& \frac{3(1-8x+20x^2)}{\sqrt{4x-1}} \cos^{-1} \left(\frac{3x-1}{2x^{3/2}}\right) \nonumber \\
&&-\frac{1-x}{2x}(2-13x+47x^2)-\frac{3}{2}(1-6x+4x^2)\log x
\end{eqnarray}
\noindent
with $x=M_Z^2/M_H^2$ \cite{Ang_dist}. The invariant 
mass $(M_*)$ spectrum of the off--shell $Z$ boson is maximal close
to the kinematical limit corresponding to zero momentum of the off-- and
on--shell $Z$ bosons in the final state:
\begin{eqnarray}
\frac{d\Gamma_H}{d M^2_*}=\frac{3 G^2_F M^4_Z\delta_Z}{16\pi^3 M_H} \,
   \frac{12 M^2_* M^2_Z+M^4_H\beta^2}{(M^2_*-M^2_Z)^2+M^2_Z\Gamma^2_Z}
   \,\beta
\end{eqnarray}
where 
$\beta$ is the $Z^*$/$Z$ three-momentum in the $H$ rest frame, in units of the
Higgs particle mass $M_H$, {\it i.e.} $\beta^2=[1-(M_Z+M_*)^2/M^2_H]
[1-(M_Z-M_*)^2/M^2_H]$. The invariant mass spectrum decreases
linearly with $\beta$ and therefore steeply with the invariant mass just 
below the threshold: 
\begin{eqnarray}
\frac{d \Gamma_H}{d M^2_*}\,\sim\,\beta\,\sim\,
                          \sqrt{\,(M_H-M_Z)^2-M^{\,2}_*\,} 
\end{eqnarray}
This steep decrease is characteristic of the decay of a scalar particle into 
two vector bosons with only two exceptions as discussed below.\s 

The second characteristic is the angular distributions of the off/on-shell 
$Z$ bosons in the final state~\cite{Ang_dist}. 
In the same notation as before,
\begin{eqnarray}
\frac{d\Gamma_H}{d\cos{\theta_1}d\cos{\theta_2}} &\sim&
   \sin^2{\theta_1} \sin^2{\theta_2} \\ 
  && +\, \frac{1}{2 \gamma^2_1\gamma^2_2 (1+\beta_1\beta_2)^2}\left[
    (1+\cos^2{\theta_1})(1+\cos^2{\theta_2}) 
   +4\, \eta_1 \eta_2
    \,\cos{\theta_1} \cos{\theta_2}\right] \nonumber
\end{eqnarray}
and
\begin{eqnarray}
\frac{d\Gamma_H}{d \varphi} \, \sim \, 1
 -\eta_1\eta_2\frac{1}{2}{\left(\frac{3\pi}{4}\right)}^2
  \frac{\gamma_1\gamma_2(1+\beta_1\beta_2)}{
   \gamma^2_1\gamma^2_2 (1+\beta_1\beta_2)^2 +2}\cos\varphi
 + \, \frac{1}{2}\frac{1}{\gamma^2_1\gamma^2_2 (1+\beta_1\beta_2)^2 
    +2}\cos2\varphi
\end{eqnarray}

\noindent
where $\beta_i, \gamma_i$ are the velocities and Lorentz-boost factors 
of the off- and on-shell $Z$ bosons, respectively. \s  

For a ${\cal CP}$ invariant theory the invariant mass and 
polar/azimuthal angular distributions can formally be written in the same
form as Eqs.(\ref{eq:ddhel1}) and (\ref{eq:ddhel2}), just modified by the 
virtual $Z^*$ propagator:
\begin{eqnarray}
\frac{d\Gamma}{d M^2_* d \cos{\theta_1} d \cos{\theta_2}}\;\; {\rm {and}} \;\; 
\frac{d\Gamma}{d M^2_* d{\varphi}}\,
   &\sim& \frac{M^2_*}{(M^2_*-M^2_Z)^2+ M^2_Z \Gamma^2_Z} \;\beta
   \, \times \, \cdots
%\,\beta\,\nonumber\\
%   &\times& \Bigg\{\,\sin^2{\theta_1}\sin^2{\theta_2}\, |{\cal T}_{00}|^2
%     +\frac{1}{2}(1+\cos^2{\theta_1})(1+\cos^2{\theta_2})
%      \left[|{\cal T}_{11}|^2+|{\cal T}_{1,-1}|^2\right]\nonumber\\
%   &&\ \ \ \ +(1+\cos^2{\theta_1})\, \sin^2{\theta_2}\,|{\cal T}_{10}|^2 
%         +\sin^2{\theta_1}\,(1+\cos^2{\theta_2})\,
%                               |{\cal T}_{01}|^2 \nonumber\\
%   &&\ \ \ \ +\frac{2v_1a_1}{v^2_1+a^2_1}\frac{2v_2a_2}{v^2_2+a^2_2}
%          \,2\, \cos{\theta_1} \cos{\theta_2}
%             \left[|{\cal T}_{11}|^2-|{\cal T}_{1,-1}|^2\right]
%	     \Bigg\}
\label{eq:ddhelx} 
\end{eqnarray}

The helicity amplitudes of the decay $H\rightarrow Z^{(*)}Z$ in the Standard
Model are given by
\begin{eqnarray}
{\cal T}_{00} = \frac{M^2_H-M^2_Z-M^2_*}{2 M_Z M_*}, \quad
{\cal T}_{11} = -1, \quad
{\cal T}_{10}={\cal T}_{01}={\cal T}_{1,-1}\, =\, 0  
\end{eqnarray}
The general helicity amplitudes are restricted by the normality
condition (\ref{eq:norm}), but not by the Bose symmetry relation
anymore. \s

The leading $\beta$ dependence of the helicity amplitudes can be 
determined by counting the number of momenta in each term of the tensor 
$T_{\mu \nu \beta_1 ... \beta_{\cal J}}$. Each momentum contracted with 
the $Z$-boson polarization vector or the $H$ polarization tensor will 
necessarily give zero or one power of $\beta$.
Furthermore, any momentum contracted with the lepton current will also
give rise to one power of $\beta$ due to the transversality of the
current. The overall $\beta$ dependence of the invariant mass spectrum can be 
derived from the squared $\beta$ dependence of the helicity amplitude 
multiplied by a single factor $\beta$ from the phase space. \s

\noindent
\underline{\bf {Odd normality:}}\\
\noindent
For the same arguments as before, the states of odd normality
${\cal J^P}=0^-$, $1^+$, $2^-$, \ldots,
can be excluded if a non-zero $\sim \, \sin^2\theta_1\,\sin^2\theta_2$
correlation has been established experimentally. Equivalently, the high
power suppression of the virtual mass distributions near the threshold
rules out all spin~$\geq 2$ states; the state ${\cal J} = 1$ can be
eliminated by non-observation of $\sim \, (1+\cos^2\theta_1)\,\sin^2 \theta_2$
and $\sin^2\theta_1\, (1+\cos^2\theta_2)$ correlations. \s

\noindent
\underline{\bf {Even normality:}}\\
\noindent
Below the threshold of two real $Z$ bosons, the states 
of even normality ${\cal J}^{\cal P} = 1^-$, $2^+$, $3^-$ \ldots. 
can be excluded by measuring the threshold behaviour of the
invariant mass spectrum and the angular correlations. \s

\noindent
{\bf Spin 1:} Every term in $T_{\mu\nu\beta}$  
  must involve at least one power of momentum so that every helicity amplitude 
  vanishes near threshold  linearly in $\beta$. As a result,
  the invariant mass spectrum decreases $\sim\beta^3$, distinct from the 
  Standard Model. \s

\begin{figure}[htb!]
\begin{center}
\includegraphics[clip=true,trim=5 5 5 5,width=11cm]{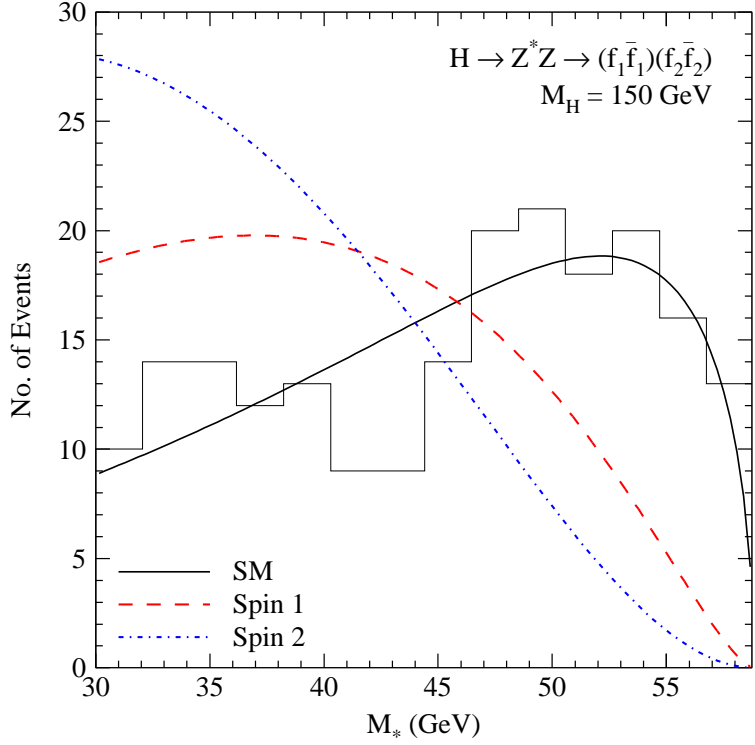}
\end{center}
\caption{{\it The threshold behaviour of the differential
distribution $d\Gamma/dM_*$ for the Standard Model and two possible
examples of spin-1 [$b_1=1/M_H$, $b_2=1/M_H^3$, $b_3=1/M_H$ and
\mbox{$b_4=1/M_H$]} and spin-2 [$c_1=0$, $c_2=1/M_H^2$, $c_3=1/M_H^2$,
$c_4=1/M_H^2$ and $c_5=1/M_H^4$] even normality bosons, with a Higgs
boson mass of 150 GeV. The histogram for the Standard Model shows the
expected result from $203$ signal events corresponding to 
an integrated luminosity of $\int {\cal L}\, dt = 300\, {\rm fb}^{-1}$ at LHC
[with efficiencies and cuts included according to the experimental
simulation Ref.\cite{Hohl}]. The curves show the exact theoretical
dependences for such scenarios, appropriately normalised.}}
\label{fig:203} 
\end{figure}

\noindent
The size of the effect is illustrated in Fig.\ref{fig:203} for a
standard sample of events at LHC for a Higgs mass $M_H = 150$\,GeV,
cf.  Ref.\cite{Hohl}, where the maximal event rate in the intermediate
mass range for $Z^*Z$ decays is expected. Standard cuts applied by the
LHC experiments have only little effect on the distributions.  In
particular, we have performed a Monte Carlo study which has
demonstrated that the rapidity and transverse momentum cuts typically
applied at the LHC do not lead to a systematic depletion of the large
$M_\star$ region that is crucial for spin measurements by the present
method. \s

The figure clearly illustrates the suppression 
of the invariant mass distributions near threshold for higher spin states
in stark contrast to the spin--0 case of the Standard Model. \s

\noindent
{\bf Spin 2:} The general spin--2 tensor contains a term
  with no momentum dependence,
  \begin{eqnarray}
  T^{\mu\nu\beta_1\beta_2} \,\sim\,    
    {g^{\mu{\beta_1}}} {g^{\nu{\beta_2}}} 
    + {g^{\mu{\beta_2}}} {g^{\nu{\beta_1}}} 
  \end{eqnarray}
  resulting in helicity amplitudes which do not vanish at threshold.
  This term however contributes to the helicity amplitudes ${\cal T}_{10}$ 
  and ${\cal T}_{01}$, leading to non-trivial 
  $(1+\cos^2\theta_1)\sin^2 \theta_2$ and $\sin^2\theta_1 (1+\cos^2\theta_2)$ 
  correlations which are absent in the Standard Model.  
  Therefore, if the invariant mass spectrum decreases linearly
  and if these polar--angle correlations are not observed experimentally,
  the spin--2 assignment to the state is ruled out. Without this 
  peculiar term in the spin-$2$ case, the spectrum falls off 
  $\sim \beta^5$ near threshold.\s

\noindent
{\bf Spin$_{ }$ $\geq$ 3:} Above spin--2 the number of 
  independent helicity amplitudes does not increase any more~\cite{Helicity} 
  and the most general spin-${\cal J}$ tensor $T_{\mu\nu\beta_1
  ... \beta_{\cal J}}$ is a direct product of a tensor $T_{\mu \nu \beta_i
  \beta_j}^{(2)}$ isomorphic with the spin-2 tensor and a symmetric 
  tensor built up by the momentum
  vectors \mbox{$k^{\beta_k}=(k_1-k_2)^{\beta_k}$} as required by the
  properties of the spin--${\cal J}$ wave-function $\varepsilon^{\beta_1 ...
  \beta_{\cal J}}$. Contracted with the wave--function, the extra ${\cal J}-2$ 
  momenta give rise to a leading power $\beta^{{\cal J}-2}$ in the helicity 
  amplitudes. The invariant mass spectrum therefore decreases near 
  threshold $\sim \beta^{2{\cal J}-3}$, {\it i.e.} with a power $\geq 3$, 
  in contrast to the single power of the Standard Model. \s

\section{Conclusions}

The analyses described above can be summarized in a few
characteristic points which cover the essential conclusions. \s

\underline{Above the threshold} for two real $Z$ bosons, $H \rightarrow ZZ$,
any odd--${\cal J}$ state
can be ruled out by observing non--zero $\sim \sin^2\theta_1\sin^2\theta_2$  
correlations. However, even--${\cal J}$ states $\geq 2$ may mimic the 
spin--0 case. Exclusion of these even--${\cal J}$ states requires 
the measurement of 
%the $\Theta, \Phi$ 
angular correlations of the $Z$ bosons with the initial state. 
It has been proven that the processes $gg, \gamma \gamma \rightarrow 
H\rightarrow ZZ$ are suitable for this purpose;
the angular distributions are an-isotropic for all spin states except 
spin--0. \s

\underline{Below the threshold} for two real bosons, $H \rightarrow Z^*Z$,  
the key is the threshold behaviour of the invariant mass spectrum
which is predicted to be linear in the $\beta$ for the
${\cal J}^{\cal P}=0^+$ Higgs boson within the Standard Model.  
All other ${\cal J}^{\cal P}$ assignments can be ruled out 
by the observation of a linear decrease
near the kinematical limit, if supplemented by angular
correlations in two exceptional cases, ${\cal J}^{\cal P} = 1^+$ and $2^+$,
{\it i.e.} observation of the $\sim \sin^2\theta_1\sin^2\theta_2$ correlation
but absence of the $(1+\cos^2\theta_1)\sin^2 \theta_2$ correlation (and 
$\sin^2\theta_1(1+\cos^2\theta_2)$). \s

The rules can be supplemented by observations specific to two cases. 
By observing non--zero $H\gamma\gamma$ and $Hgg$ couplings, the ${\cal J} 
= 1$ assignment can elegantly be ruled out by Yang's theorem in particular, 
and for all odd spins in general \cite{Jacob}. \s

The above formalism can be generalized easily to rule out mixed
normality states with spin~$\geq 1$. For a Higgs boson of mixed
normality we cannot use Eq.(\ref{eq:norm}) anymore to derive the
simple form of the differential decay width in Eqs.(\ref{eq:ddhel1}) 
and (\ref{eq:ddhel2}). In particular, the double polar--angle
distribution, Eq.(\ref{eq:ddhel1}), is modified to include 
linear terms proportional to $\cos\theta_1$ or $\cos\theta_2$, indicative of 
${\cal CP}$ violation~\cite{haggun}. The analysis for identifying the
spin of the Higgs particle, however, proceeds exactly as before in
the fixed normality case, since the most general vertex will be
the sum of the even and odd normality tensors. \s

\vspace {8mm}
\section{Appendix}
\noindent
{\bf (a)} In the Standard Model the general combined polar and azimuthal 
correlation is given by the expression
\begin{eqnarray}
\frac{d\Gamma_H}{d\cos{\theta_1} d\cos{\theta_2} d \varphi}& \sim&
   \sin^2{\theta_1} \sin^2{\theta_2} - \frac{1}{2 \gamma^2(1+\beta^2)} 
   \sin{2\theta_1} \sin{2\theta_2} \cos{\varphi} \nonumber \\
&& \hspace{-1.5cm} 
   +\, \frac{1}{2 \gamma^4(1+\beta^2)^2}
    \left[(1+\cos^2{\theta_1})(1+\cos^2{\theta_2})
   +\sin^2{\theta_1} \sin^2{\theta_2} \cos{2\varphi} \right] \nonumber \\
&& \hspace{-1.5cm} 
   +\, \eta_1 \eta_2 \,
    \frac{2}{\gamma^2(1+\beta^2)} \left[-\sin{\theta_1}\sin{\theta_2}
    \cos{\varphi}
   + \frac{1}{\gamma^2(1+\beta^2)} \cos{\theta_1} \cos{\theta_2} \right]
\end{eqnarray}
\s
\newpage
\noindent
{\bf (b)} while in the general ${\cal CP}$ conserving case
\vspace{3mm} 
\begin{eqnarray}
\frac{d\Gamma}{d\cos\theta_1 d\cos\theta_2 d\varphi} &\sim&
 \sin^2{\theta_1}\sin^2{\theta_2}\, |{\cal T}_{00}|^2
      +\frac{1}{2}(1+\cos^2{\theta_1})(1+\cos^2{\theta_2})
        \left[|{\cal T}_{11}|^2+|{\cal T}_{1,-1}|^2\right]\nonumber\\
     &&{ }\hskip -3.3cm +(1+\cos^2{\theta_1})\, \sin^2{\theta_2}\,
                |{\cal T}_{10}|^2
	       +\sin^2{\theta_1}\,(1+\cos^2{\theta_2})\,|{\cal T}_{01}|^2 \nonumber\\[2mm]
     &&{ }\hskip -3.3cm  + 2\, \eta_1 \eta_2\, \cos{\theta_1} \cos{\theta_2}
	         \left[|{\cal T}_{11}|^2-|{\cal T}_{1,-1}|^2\right]
	       \nonumber\\[2mm]
     &&{ }\hskip -3.3cm  + 2 \sin\theta_1\sin\theta_2
          \left[\cos\theta_1\cos\theta_2
                \real({\cal T}_{11}{\cal T}^*_{00}
		     -{\cal T}_{10}{\cal T}^*_{0,-1})
		+\eta_1\eta_2
                \real({\cal T}_{11}{\cal T}^*_{00}
		     +{\cal T}_{10}{\cal T}^*_{0,-1})\right]\cos\varphi
	       \nonumber\\[2mm]
     &&{ }\hskip -3.3cm - 2 \sin\theta_1\sin\theta_2
          \left[\eta_2\cos\theta_1
                \imag({\cal T}_{11}{\cal T}^*_{00}
		     +{\cal T}_{10}{\cal T}^*_{0,-1})
		+\eta_1\cos\theta_2
                \imag({\cal T}_{11}{\cal T}^*_{00}
		     -{\cal T}_{10}{\cal T}^*_{0,-1})\right]\sin\varphi
	       \nonumber\\[2mm]
     &&{ }\hskip -3.3cm +\frac{1}{2}\sin^2{\theta_1}\sin^2{\theta_2}
	\real({\cal T}_{11}{\cal T}^*_{-1,-1})
                \cos 2\varphi
\label{eq:ddhel3}
\end{eqnarray}
using the same notation as before.
\s

\section*{Acknowledgments}

We are very grateful to Fabiola Gianotti for encouraging discussions
and, in particular, the careful reading of the manuscript.
The work is supported in part by the European Union (HPRN-CT-2000-00149) 
and by the Korean Research Foundation (KRF-2000-015-050009).

\end{document}